\documentclass[prd,showpacs,preprint]{revtex4}
\usepackage{amssymb}
\usepackage{amsmath}
\usepackage{graphicx}
\usepackage{multirow}

\begin{document}

\title{Next-to-leading order QCD corrections to a heavy resonance production
and decay into top quark pair at the LHC}
\author{Jun Gao}
\author{Chong Sheng Li}\email{csli@pku.edu.cn}
\author{Bo Hua Li}
\author{Hua Xing Zhu}
\affiliation{Department of Physics and State Key Laboratory of
Nuclear Physics and Technology, Peking University, Beijing 100871,
China}
\author{C.-P.Yuan}\email{yuan@pa.msu.edu}
\affiliation{Department of Physics and Astronomy, Michigan State
University, East Lansing, 48824, USA}

\date{\today}

\pacs{12.38.Bx,~12.60.-i,~14.65.Ha}

\begin{abstract}
We present a complete next-to-leading order (NLO) QCD calculation to
a heavy resonance production and decay into a top quark pair at the
LHC, where the resonance could be either a Randall-Sundrum (RS)
Kaluza-Klein (KK) graviton $G$ or an extra gauge boson $Z'$. The
complete NLO QCD corrections can enhance the total cross sections by
about $80\%- 100\%$ and $20\%- 40\%$ for the $G$ and the $Z'$,
respectively, depending on the resonance mass. We also explore in
detail the NLO corrections to the polar angle distributions of the
top quark, and our results show that the shapes of the NLO
distributions can be different from the leading order (LO) ones for
the KK graviton. Moreover, we study the NLO corrections to the spin
correlations of the top quark pair production via the above process,
and find that the corrections are small.
\end{abstract}

\maketitle

\section{introduction}\label{s1}
The top quark is the heaviest particle so far discovered, with a mass
close to the electroweak symmetry breaking scale, and closely related
to various new physics models beyond the standard model (SM). Thus it
provides an effective probe for the electroweak symmetry breaking
mechanism and the new physics beyond the SM through studying its
production and decay at colliders. The Large Hadron Collider (LHC) is
running now with a center of mass energy $\sqrt s =7{\rm\ TeV}$, and
will collect 1 $\rm fb^{-1}$ experimental data during the initial
run. After this initial state the LHC will turn to $\sqrt s =14{\rm \
TeV}$, with a design luminosity of $\sim 10\ {\rm fb^{-1}/yr}$ there
will be $8\times 10^6$ top quark pairs and $3\times 10^6$ single top
quarks produced yearly. As a result of all these, the precision
measurement of the top quark properties, such as the mass, the
production cross sections, the kinematic distributions, and the spin
correlation effects, will be one of the prime tasks in the
experiments at the LHC, and any deviations from the SM predictions
will definitely be a hint for new physics beyond the SM.

To explore the connections between the new physics and the top quark,
one possibility is to study the top quark pair invariant mass
distribution and look for possible resonances since many new physics
models predict the existence of a new resonance with a mass around
${\rm TeV}$, which can decay into a top quark pair, such as the
Technicolor~\cite{Hill:2002ap}, Topcolor~\cite{Hill:1991at}, Little
Higgs~\cite{ArkaniHamed:2001nc}, general $Z'$
models~\cite{Langacker:2008yv,Godfrey:2008vf}, and Randall-Sundrum
(RS) models~\cite{Randall:1999ee}. In addition, in many such models
the interaction between the heavy resonance and the top quark is
enhanced as compared to the other fermions and the resonance will
mainly decay into a top quark pair, for example, the Kaluza-Klein
(KK) excitations of the graviton~\cite{Fitzpatrick:2007qr}, the
gluon~\cite{Gherghetta:2000qt} as well as the weak gauge
bosons~\cite{Agashe:2007ki} in the extended RS models. Once we have
discovered such a resonance in the top quark pair invariant mass
distribution, the next step is to measure its spin and couplings, and
finally determine the underlying new physics dynamics, which have
been studied in Refs.~\cite{Barger:2006hm,Frederix:2007gi}. It has
been suggested in Refs.~\cite{Barger:2006hm,Frederix:2007gi} that it
is possible to extract the spin and coupling information of the
resonance from the top quark polar angle distributions and the spin
correlations of the top quark pair. Those studies were carried out at
the leading-order (LO) in QCD interactions. However, the
next-to-leading order (NLO) QCD corrections may be large, for
example, the NLO QCD corrections can enhance the cross sections of
the single RS KK graviton or the $Z'$ production by about
$70\%$~\cite{Mathews:2005bw,Li:2006yv} and $20\%$~\cite{Ball:2007zza}
respectively, so it is necessary to examine whether the QCD
corrections will change the tree-level results and some conclusions
of Refs.~\cite{Barger:2006hm,Frederix:2007gi} or not. In this paper
we investigate the NLO QCD effects to a heavy resonance production
and decay into a top quark pair, i.e., $pp\rightarrow X(color\
singlet)\rightarrow t\bar t$, at the LHC, where the $SU(3)_C$ color
singlet state $X$ could be either a RS KK graviton $G$ or an extra
gauge boson $Z'$.

The arrangement of this paper is as follows. Section~\ref{s2} is a
brief review to the relevant models. In Sec.~\ref{s3} we show the
details of the NLO calculations. Section~\ref{s4} contains the
numerical results, and Sec.~\ref{s5} is a brief summary. The Appendix
collects some analytic results at the LO.

\section{the models}\label{s2}
\subsection{The RS KK graviton}
In the RS model, a single extra dimension is compactified on a $\rm
S^1/Z_2$ orbifold with a radius $r$, which is not too large as compared
to the Planck length. Two 3-branes, the Planck brane and the TeV
brane, are located at the orbifold fixed points $\phi=0,\pi$,
respectively, and the spacetime between the two 3-branes is simply a
slice of a five-dimensional anti-de Sitter geometry. The
five-dimensional warped metric is given by
\begin{equation}
ds^2=e^{-2kr|\phi|}\eta_{\mu \nu}dx^{\mu}dx^{\nu}-r^2d\phi^2,
\end{equation}
where $\phi$ is the five-dimensional coordinate, and $k \sim M_P$ is
the curvature scale. By requiring $kr\sim 12$, one can suppress the
Planck scale to $ M_P e^{-k\pi r}\sim O({\rm TeV})$ on the TeV brane,
and then solve the gauge hierarchy problem. The gravity fields are
treated as fluctuations under the background metric, and after
expanding the gravity fields in the extra dimension we get infinite
massive KK gravitons, which can interact with the SM
fields~\cite{Csaki:2004ay}.

\begin{figure}[h!]
\includegraphics[width=0.7\textwidth]{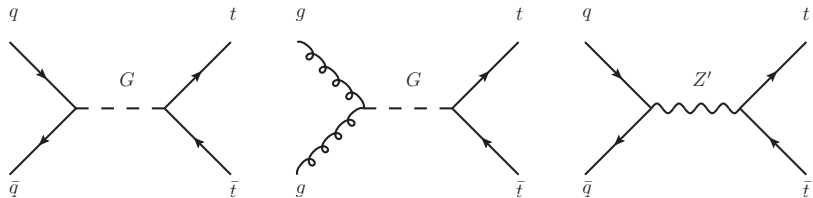}
\caption[]{Tree-level Feynman diagrams for the heavy resonance
production and decay into a top quark pair.} \label{f0}
\end{figure}

The RS KK graviton can be produced through both the $gg$ fusion and
the $q\bar q$ annihilation at the LO as shown in Fig.~\ref{f0}. The
detailed Feynman rules of the graviton couplings can be found in
Ref.~\cite{Han:1998sg}, and the propagator of the graviton in the
unitary gauge in $n$ dimensions is given by~\cite{Mathews:2005bw}
\begin{equation}
P_G(k)={iB_{\mu\nu_,\rho\sigma}(k)\over k^2-m_X^2+im_X\Gamma_X},
\end{equation}
with
\begin{eqnarray}
B_{\mu\nu_,\rho\sigma}(k)&=&\left(g_{\mu\rho}-\frac{k_{\mu}k_{\rho}}{m_X^2}
\right)\left(g_{\nu\sigma}-\frac{k_{\nu}k_{\sigma}}{m_X^2}\right)+
\left(g_{\mu\sigma}-\frac{k_{\mu}k_{\sigma}}{m_X^2}
\right)\left(g_{\nu\rho}-\frac{k_{\nu}k_{\rho}}{m_X^2}\right)\nonumber
\\
&&-{2\over n-1}\left(g_{\mu\nu}-\frac{k_{\mu}k_{\nu}}{m_X^2}
\right)\left(g_{\rho\sigma}-\frac{k_{\rho}k_{\sigma}}{m_X^2}\right),
\end{eqnarray}
where $m_X$ and $\Gamma_X$ are the mass and the width of the heavy
resonance, respectively.

\subsection{The extra gauge boson $Z'$}
The extra gauge boson $Z'$ could arise from an additional $U(1)'$
gauge symmetry~\cite{Langacker:2008yv}. It could also be the KK
excitation of the electroweak gauge bosons. It can only be produced
through the $q\bar q$ annihilation at the LO, and its generic
couplings to quarks are as follow
\begin{equation}
Z'q\bar q \sim
\gamma_{\mu}\left(a_L\frac{1-\gamma_5}{2}+a_R\frac{1+\gamma_5}{2}\right).
\end{equation}
We considered four combinations of $a_L$ and $a_R$, i.e., the pure
vector coupling, $a_L=a_R=1$, the axial-vector coupling,
$a_L=-a_R=-1$, the right-handed coupling, $a_L=0$, $a_R=1$, and the
left-handed coupling, $a_L=1$, $a_R=0$, which are denoted by $Z'_1$,
$Z'_2$, $Z'_3$ and $Z'_4$, respectively. The propagator of the $Z'$
in the unitary gauge is given by
\begin{equation}
P_{Z'}(k)=\frac{i}{k^2-m_X^2+im_X\Gamma_X}\left(-g_{\mu\nu}+{k_{\mu}k_{\nu}\over
m_X^2}\right).
\end{equation}

In our calculations of the process $pp\rightarrow X(color\
singlet)\rightarrow t\bar t$, what we mainly concern about are the
ratios of the NLO results to the LO ones, so it is not necessary to
specify the actual values of all the couplings. We simply assume the
mass of the heavy resonance to be around ${\rm TeV}$ scale, which is
not yet excluded by the current experiments. Besides, we only
consider the narrow resonance cases and fix $\Gamma_X/m_X=1\%$ at
both the LO and the NLO. We do not expect our conclusions to be
largely modified even if $\Gamma_X/m_X$ increases to be at the order
of 10\%. Detailed discussions on the SM backgrounds and the discovery
potential of the process can be found in
Refs.~\cite{Barger:2006hm,Frederix:2007gi}.

\section{the NLO formalism}\label{s3}
The complete NLO QCD corrections to the process $pp\rightarrow
X(color\ singlet)\rightarrow t\bar t$ can be factorized into two
independent gauge invariant parts, i.e., the heavy resonance produced
at the NLO with a subsequent decay at the LO, and produced at the LO
with a subsequent decay at the NLO, similar to the cases studied in
Ref.~\cite{Campbell:2004ch}. The box diagrams, and the corresponding
real correction diagrams, that connect the initial and the final
states do not contribute to the squared matrix elements up to the NLO
as the heavy resonance is a $SU(3)_C$ color singlet particle. This
whole procedure can be illustrated as follows
\begin{eqnarray}
|{\mathcal M}^{tree}_{2\rightarrow 2}|^2&=&|{\mathcal
M}^{tree}_{pro}|^2\otimes|{\mathcal M}^{tree}_{dec}|^2\otimes|P_X|^2, \nonumber \\
|{\mathcal M}^{real}_{2\rightarrow 3}|^2&=&\left\{|{\mathcal
M}^{tree}_{pro}|^2\otimes|{\mathcal M}^{real}_{dec}|^2+|{\mathcal
M}^{real}_{pro}|^2\otimes|{\mathcal M}^{tree}_{dec}|^2\right\}\otimes|P_X|^2, \nonumber \\
{\mathcal M}^{tree*}_{2\rightarrow 2}{\mathcal
M}^{loop}_{2\rightarrow 2}&=&\left\{|{\mathcal
M}^{tree}_{pro}|^2\otimes ({\mathcal M}^{tree*}_{dec}{\mathcal
M}^{loop}_{dec})+|{\mathcal M}^{tree}_{dec}|^2\otimes ({\mathcal
M}^{tree*}_{pro}{\mathcal M}^{loop}_{pro})\right\}\otimes|P_X|^2,
\end{eqnarray}
we have suppressed the possible Lorentz indices here for simplicity.

\begin{figure}[h!]
\includegraphics[width=0.7\textwidth]{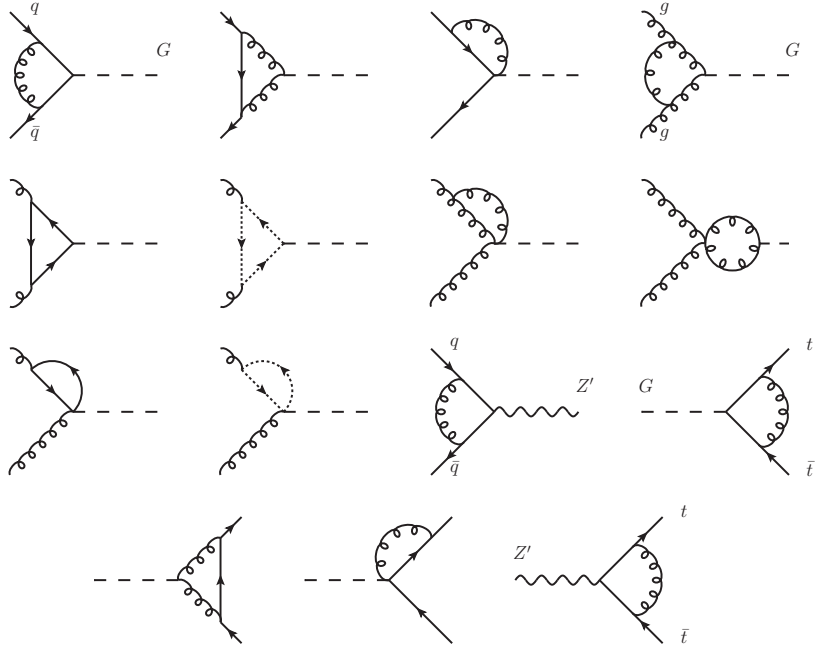}
\caption[]{Some one-loop Feynman diagrams for the production and the
decay of the heavy resonances. Others not shown can be obtained by
the exchange of the external quark or gluon lines.} \label{f1}
\end{figure}

We calculate the full squared matrix elements using the propagators
($P_X$) of the heavy resonance given in Sec.~\ref{s2}, which can
incorporate the full spin correlations between the production and
decay processes in order to generate the correct kinematic
distributions. We carry out all the QCD calculations in the 't
Hooft-Feynman gauge and use the dimension regularization
scheme~\cite{'tHooft:1972fi} (with the naive $\gamma_5$
prescription~\cite{Chanowitz:1979zu}) in $n=4-2\epsilon$ dimensions
to regularize all the divergences. The one-loop Feynman diagrams for
the production and the decay of the heavy resonance are shown in
Fig.~\ref{f1}.

\begin{figure}[h!]
\includegraphics[width=0.7\textwidth]{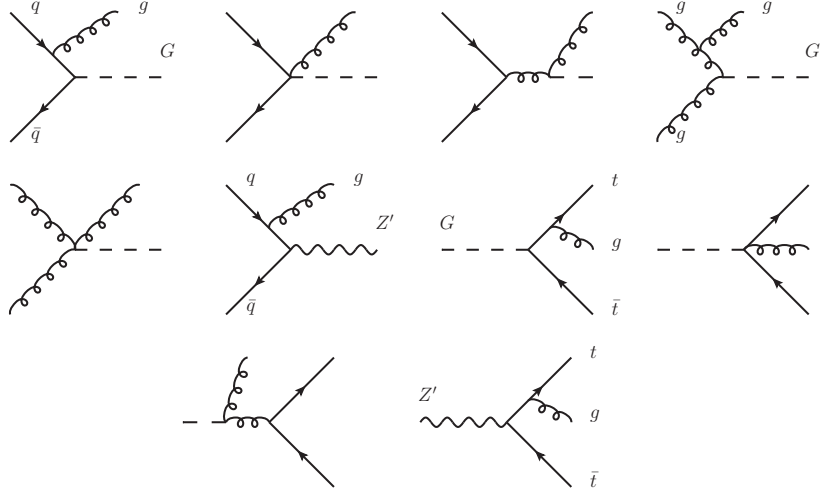}
\caption[]{Some real correction Feynman diagrams for the production
and the decay of the heavy resonances. Others not shown can be
obtained by the exchange of the external quark or gluon lines.}
\label{f2}
\end{figure}

Fig.~\ref{f2} shows the real correction Feynman diagrams for the
production and the decay of the heavy resonance. The infrared
divergences of the real corrections are extracted by using the two
cutoff phase space slicing method~\cite{Harris:2001sx}. Due to the
limited space here we do not reproduce the details of the method.

\section{Numerical results}\label{s4}
In our numerical calculations we choose the input parameters to be
$m_{top}=171{\rm\ GeV}$, $m_Z=91.188{\rm\ GeV}$, and
$\alpha_s(m_Z)=0.118$~\cite{Amsler:2008zzb}. The running QCD coupling
constant is evaluated at the three-loop order~\cite{Amsler:2008zzb}
with $n_f=5$, and the CTEQ6M (CTEQ6L1) parton distribution function
(PDF) set~\cite{Pumplin:2002vw} is used through the NLO (LO)
calculations. We set both the renormalization and factorization
scales equal to the mass of the heavy resonance, unless specified.
Besides, in the two cutoff phase slicing method there are two
arbitrary cutoff parameters, i.e., the soft cutoff $\delta_s$ and the
collinear cutoff $\delta_c$. We have checked the cutoff dependence of
all our numerical results, and found that the dependence is
negligibly small for $\delta_s\leq1\times 10^{-3}$, so we choose
$\delta_s=1\times10^{-3}$ and $\delta_c=\delta_s/50$ to obtain the
numerical results presented below.

\subsection{The total cross sections}\label{tot}
\begin{figure}[h!]
\includegraphics[width=0.4\textwidth]{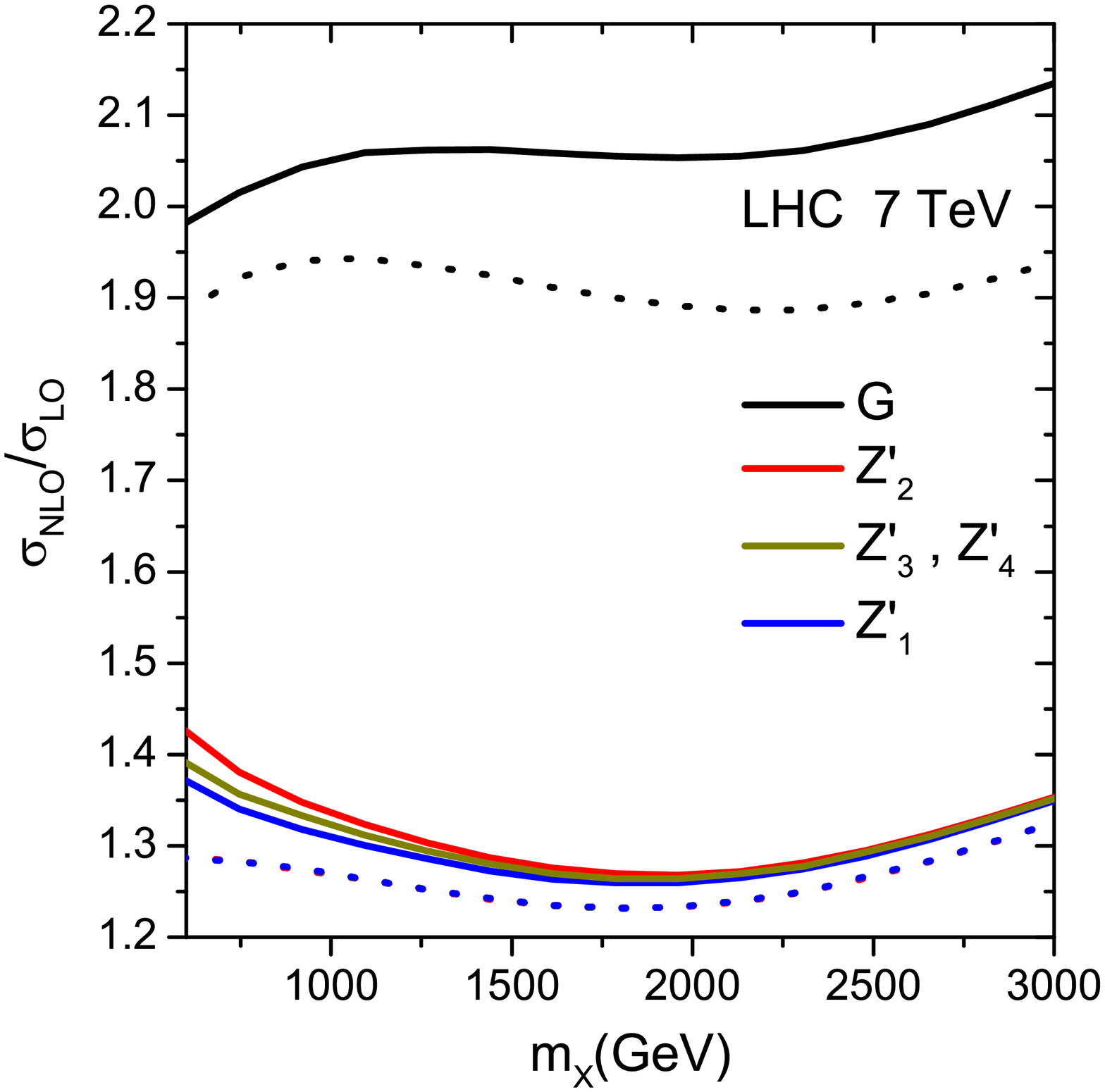}
\includegraphics[width=0.4\textwidth]{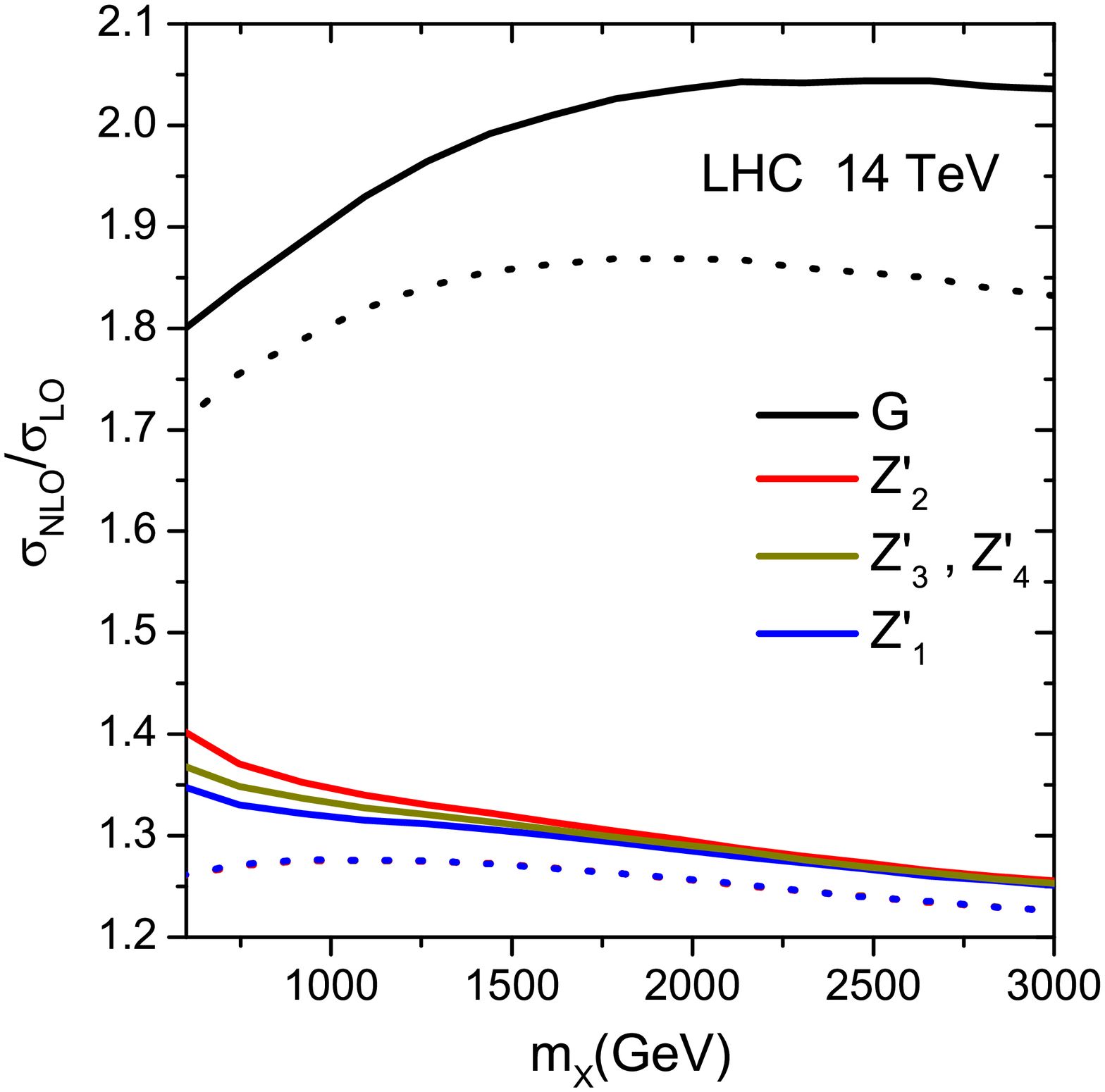}
\caption[]{The NLO K factors as functions of the heavy resonance mass
at the LHC, the solid and dotted lines correspond to including the
total QCD corrections and the QCD corrections from the production
part alone, respectively. The four groups of the curves from the top
to the bottom correspond to the $G$, $Z'_2$, $Z'_3(Z'_4)$, and
$Z'_1$, respectively.} \label{f3}
\end{figure}

\begin{figure}[h!]
\includegraphics[width=0.4\textwidth]{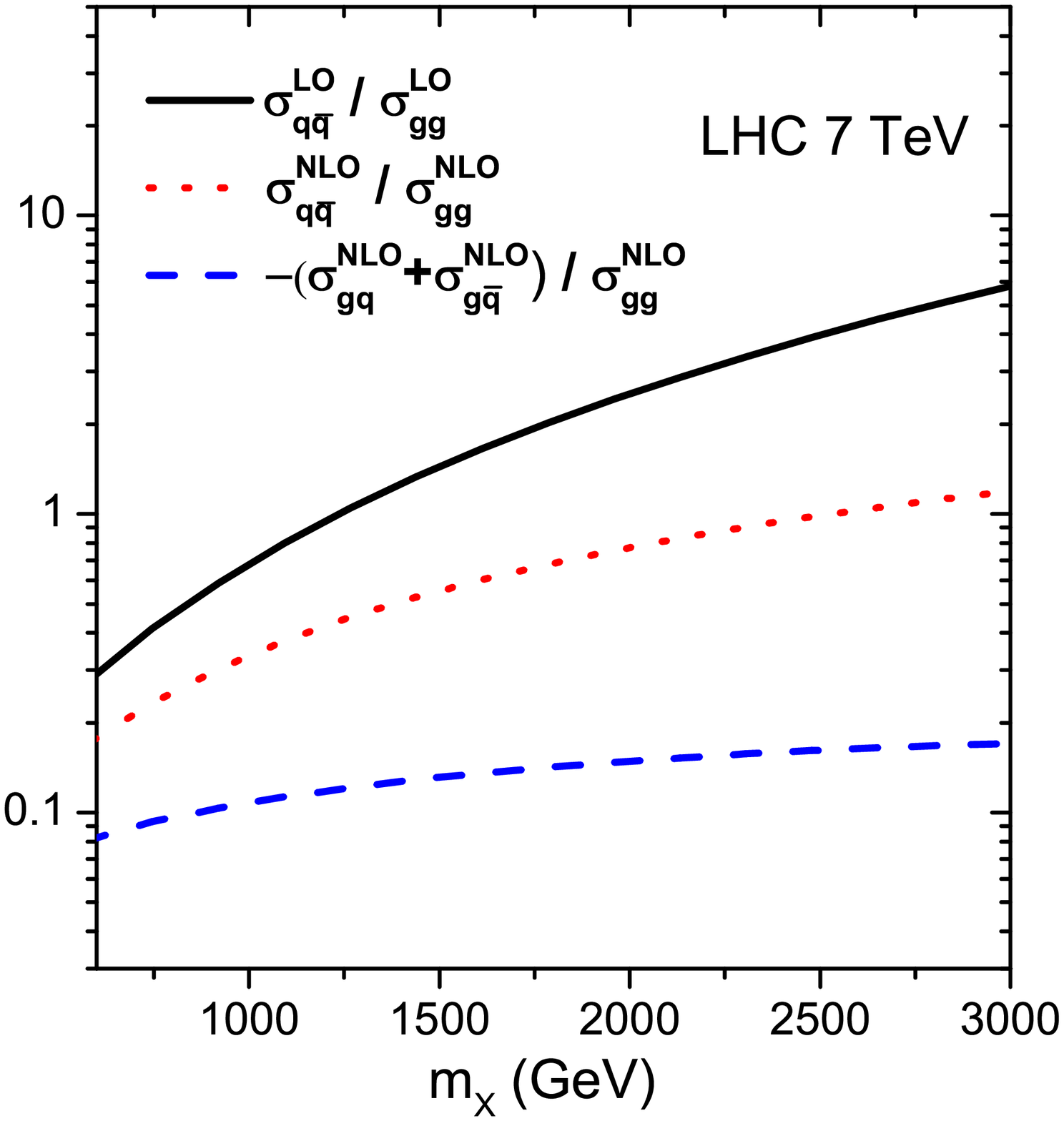}
\includegraphics[width=0.4\textwidth]{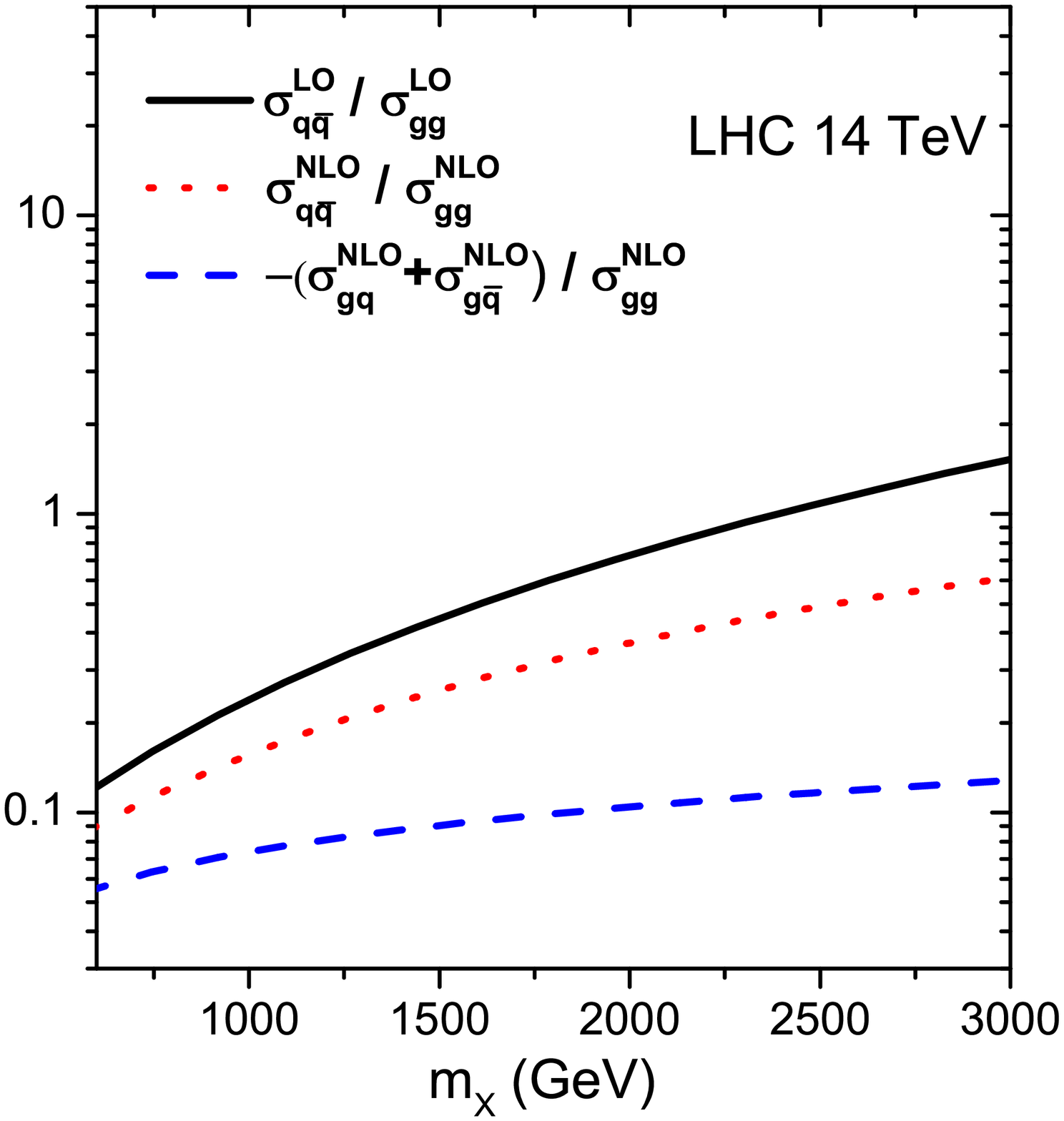}
\caption[]{The ratios of the total cross sections from different
channels for the graviton as functions of the graviton mass at both
the LO and the NLO.}
\label{f4}
\end{figure}

In Fig.~\ref{f3} we show the NLO K factor, defined as the ratio of
the NLO cross section $\sigma_{NLO}$ to the LO cross section
$\sigma_{LO}$, as a function of the heavy resonance mass at the LHC
with different center of mass energies. We can see that the total NLO
QCD corrections can be large, which can enhance the total cross
sections by about $80\%- 100\%$ and $20\%- 40\%$ for the $G$ and all
four types of $Z'$ bosons, respectively, depending on the resonance
mass. The NLO corrections from the production part are dominant,
while the ones from the decay part are relatively small, but can
still reach above ten percent in some regions. Our results of the NLO
K factors of the production part agree with the ones given in
Refs.~\cite{Li:2006yv,Ball:2007zza}, where the total cross sections
have been summed over the spins of the heavy resonance directly. In
the following parts of our paper we will only show the results of the
total NLO QCD corrections for simplicity. We further present the
ratios of the total cross sections from the different channels for
the graviton at both the LO and the NLO in Fig.~\ref{f4}. It can be
seen that the contribution from the $gg$ channel is dominant at the
low $m_X$ value region due to the large PDF of the gluon, and the
contribution from the $q\bar q$ channel becomes important at the high
$m_X$ value region since the PDF of the valence quark decreases more
slowly than the gluon. And the NLO corrections can change the ratio
of the contribution from the $q\bar q$ channel to the one from the
$gg$ channel significantly.

\subsection{The polar angle and invariant mass distributions}
It has been shown in Refs.~\cite{Barger:2006hm,Frederix:2007gi} that
the polar angle distributions of the top quark are the key points to
extract the spin and coupling information of the heavy resonance.
The definition of this polar angle depends on the reference frame and
axis chosen, here we considered two kinds of polar angles, one is the
Collins-Soper angle $\theta_S$~\cite{Collins:1977iv}, which is
defined to be the angle between the top quark momentum and the axis
that bisects the angle between the momentums of the incoming hadrons
($\vec{p}_A$ and $-\vec{p}_B$) in the $t\bar t$ rest frame, and for
the $Z'$ we can define $\theta^*$ as the angle in the $t\bar t$ rest
frame between the top quark momentum and the incident quark momentum
which can be determined by the longitudinal boost direction of the
$t\bar t$ rest frame at the LHC~\cite{Barger:2006hm}.

\begin{figure}[h!]
\includegraphics[width=0.4\textwidth]{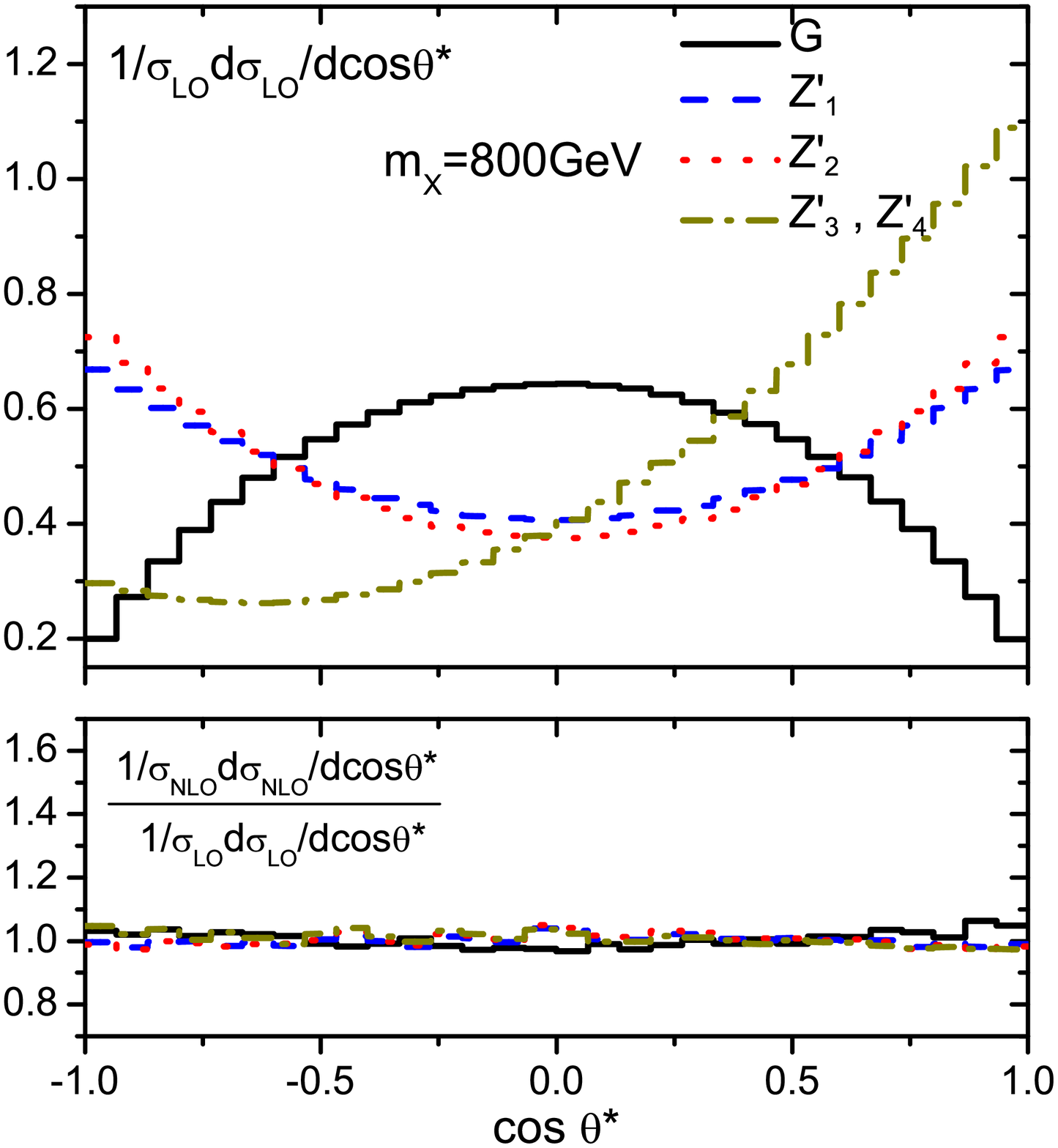}
\includegraphics[width=0.4\textwidth]{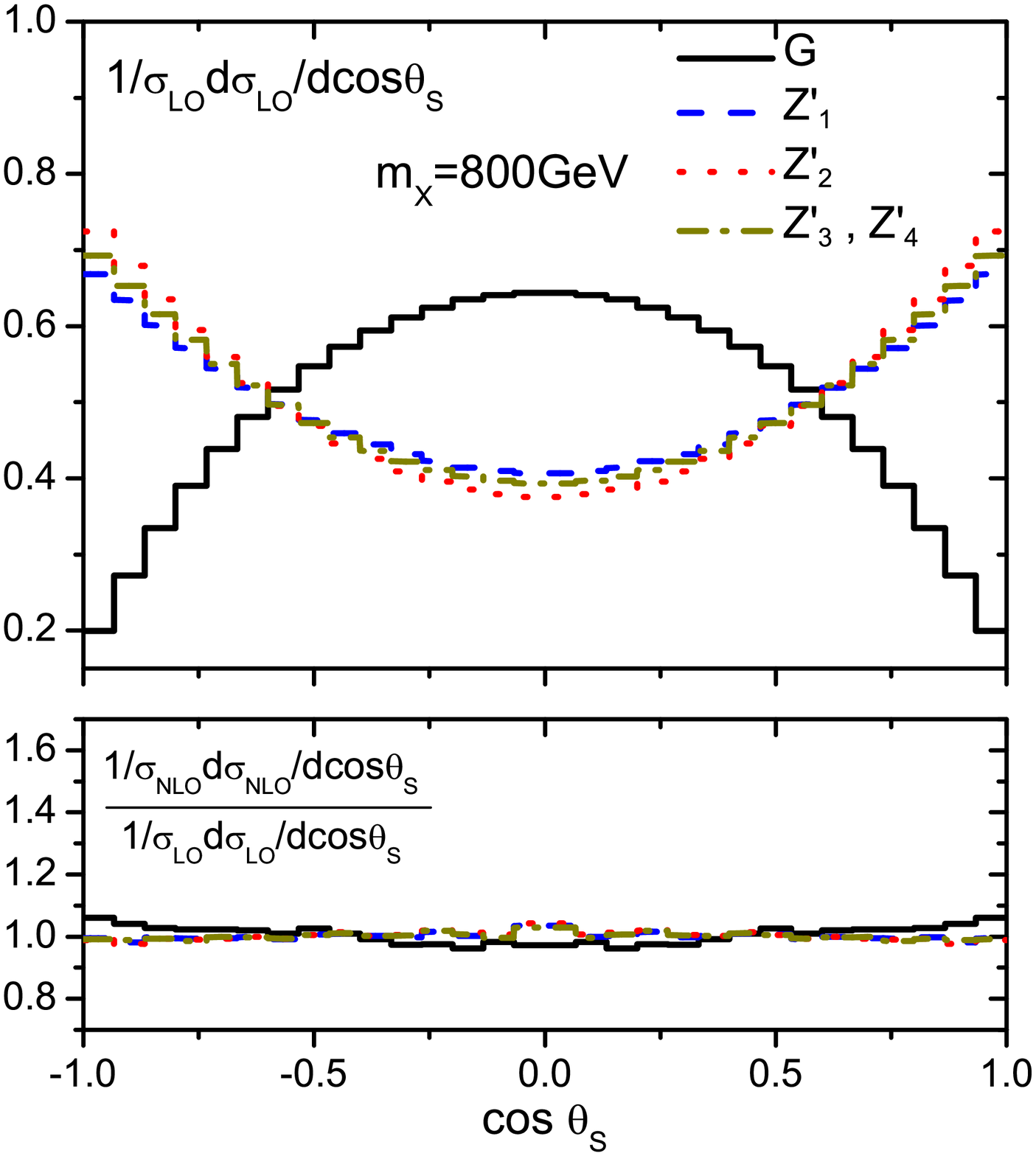}
\caption[]{The normalized top quark polar angle distributions at the LO and the
ratios of the normalized NLO distributions to the LO ones, at the LHC
with $\sqrt s=14{\rm\ TeV}$ for $m_X=800{\rm\ GeV}$.}
\label{f5}
\end{figure}

\begin{figure}[h!]
\includegraphics[width=0.4\textwidth]{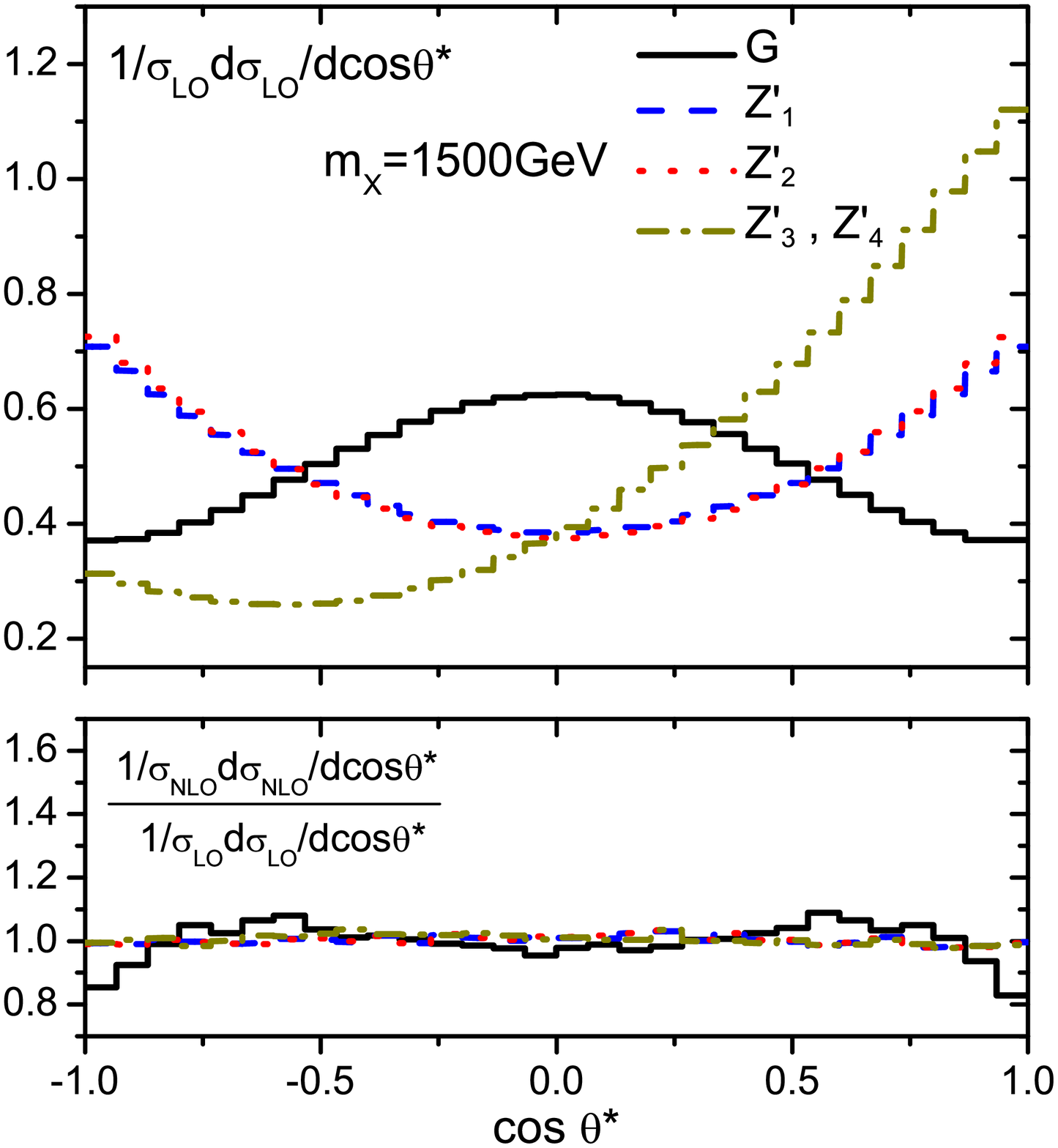}
\includegraphics[width=0.4\textwidth]{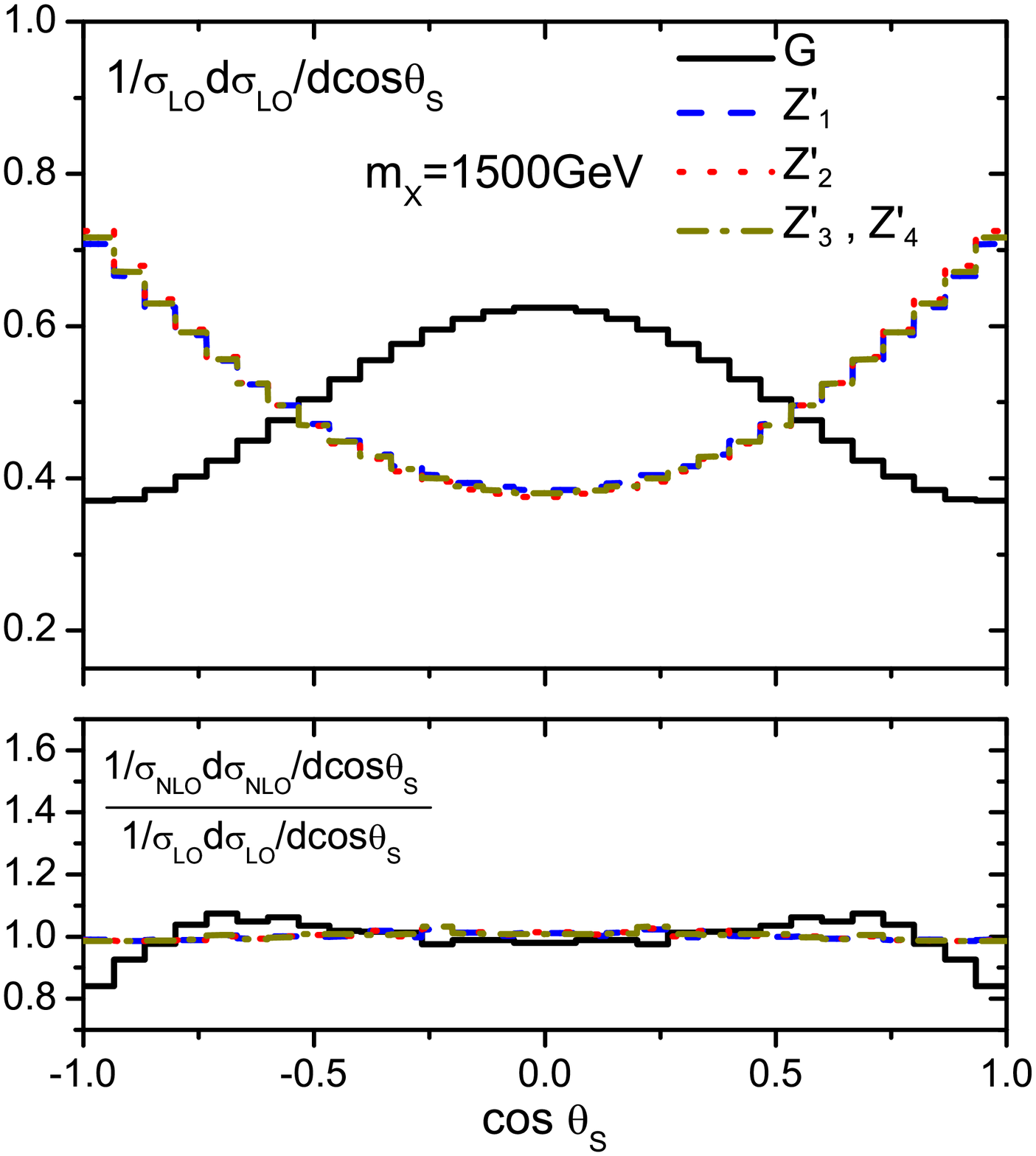}
\caption[]{The normalized top quark polar angle distributions at the LO and the
ratios of the normalized NLO distributions to the LO ones, at the LHC
with $\sqrt s=14{\rm\ TeV}$ for $m_X=1500{\rm\ GeV}$.}
\label{f6}
\end{figure}

\begin{figure}[h!]
\includegraphics[width=0.4\textwidth]{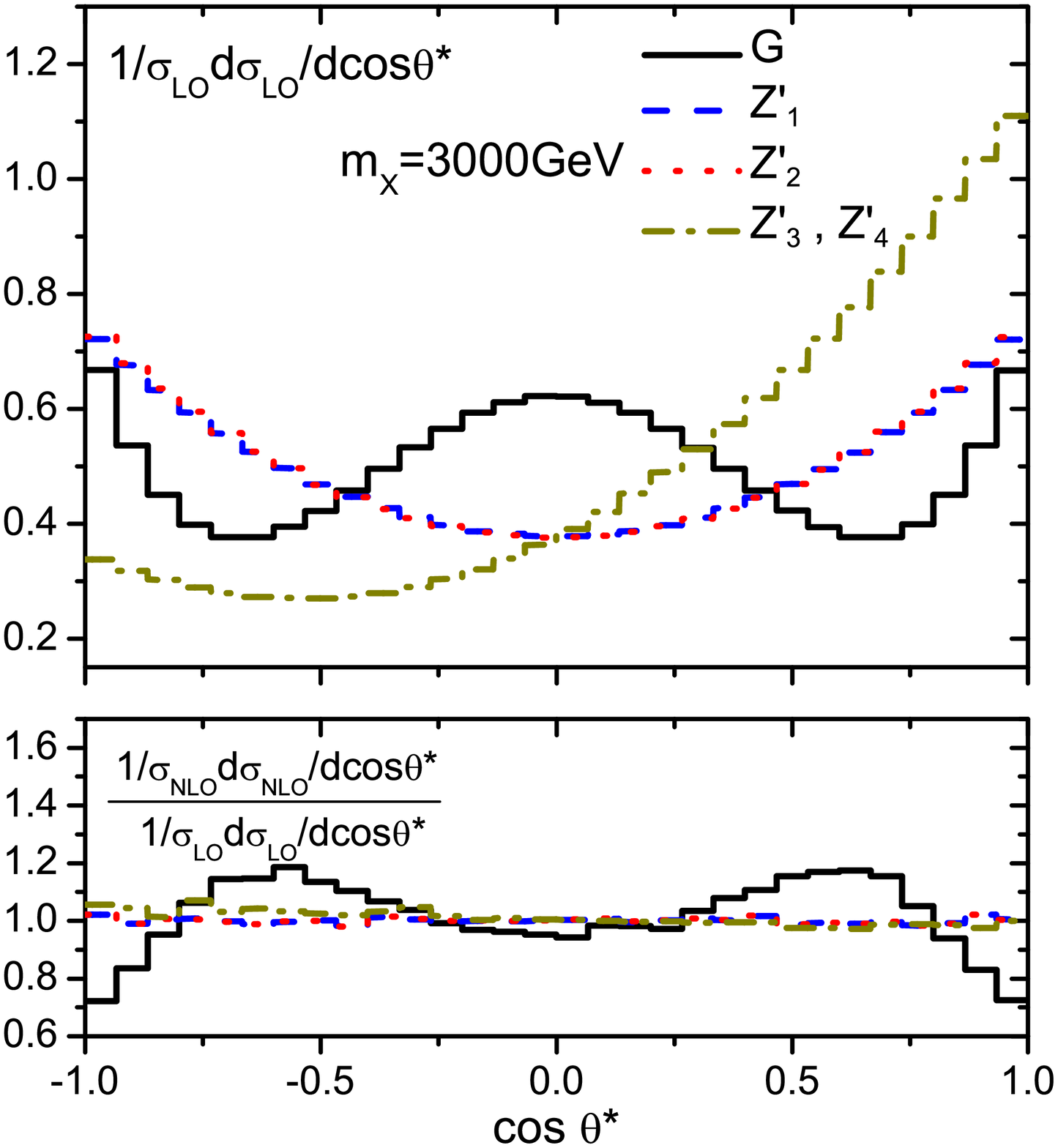}
\includegraphics[width=0.4\textwidth]{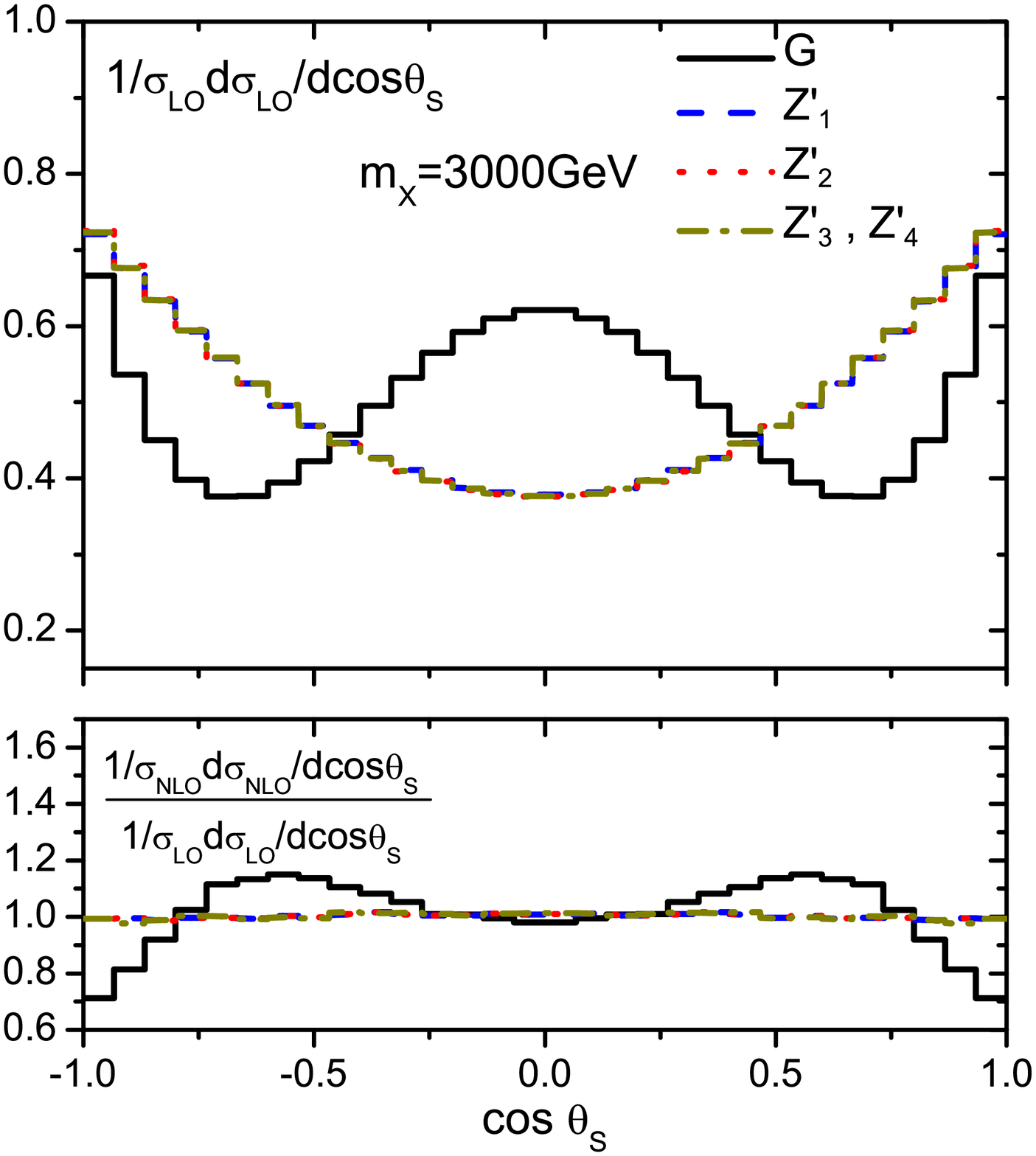}
\caption[]{The normalized top quark polar angle distributions at the LO and the
ratios of the normalized NLO distributions to the LO ones, at the LHC
with $\sqrt s=14{\rm\ TeV}$ for $m_X=3000{\rm\ GeV}$.}
\label{f7}
\end{figure}

\begin{figure}[h!]
\includegraphics[width=0.4\textwidth]{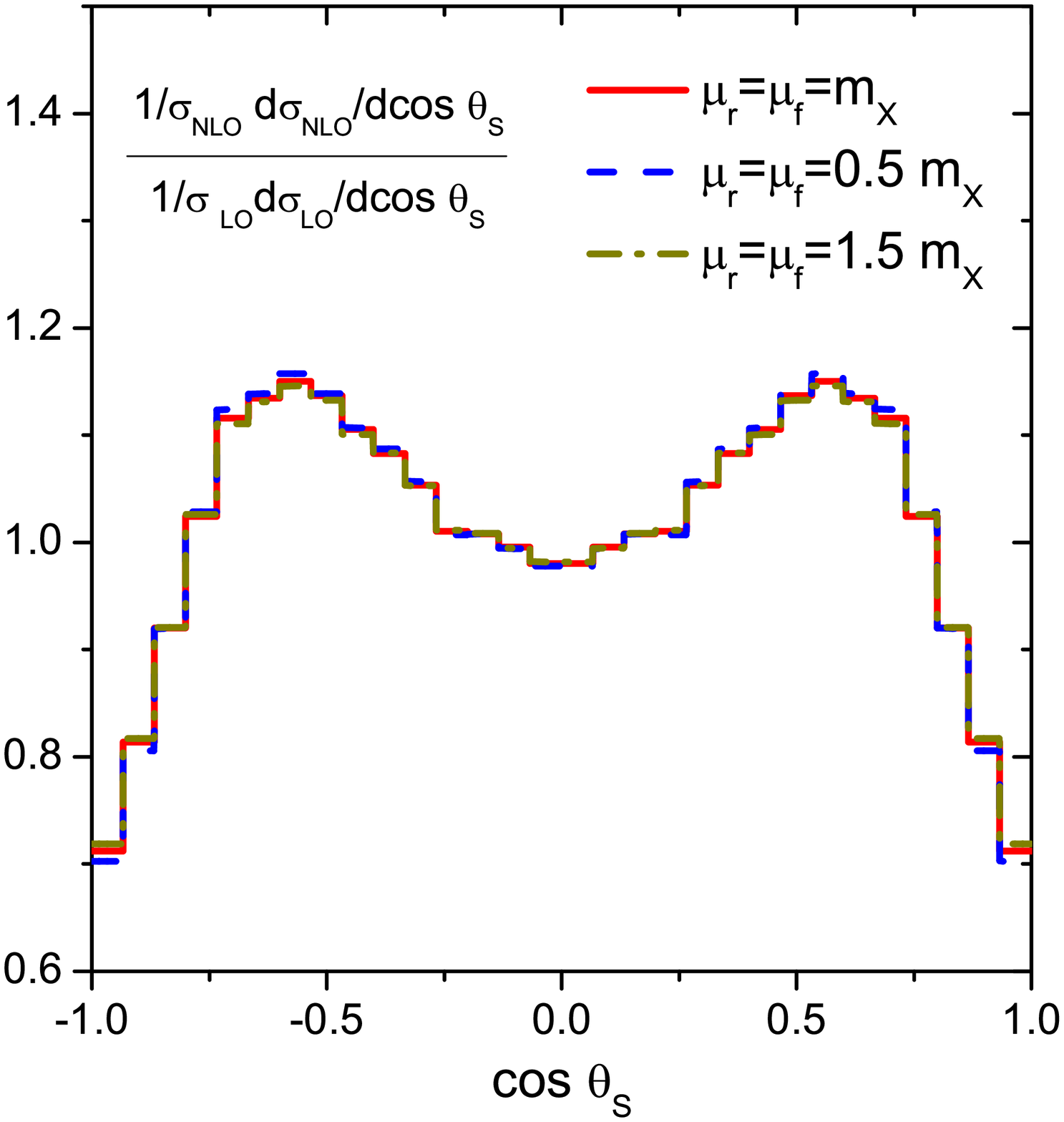}
\includegraphics[width=0.4\textwidth]{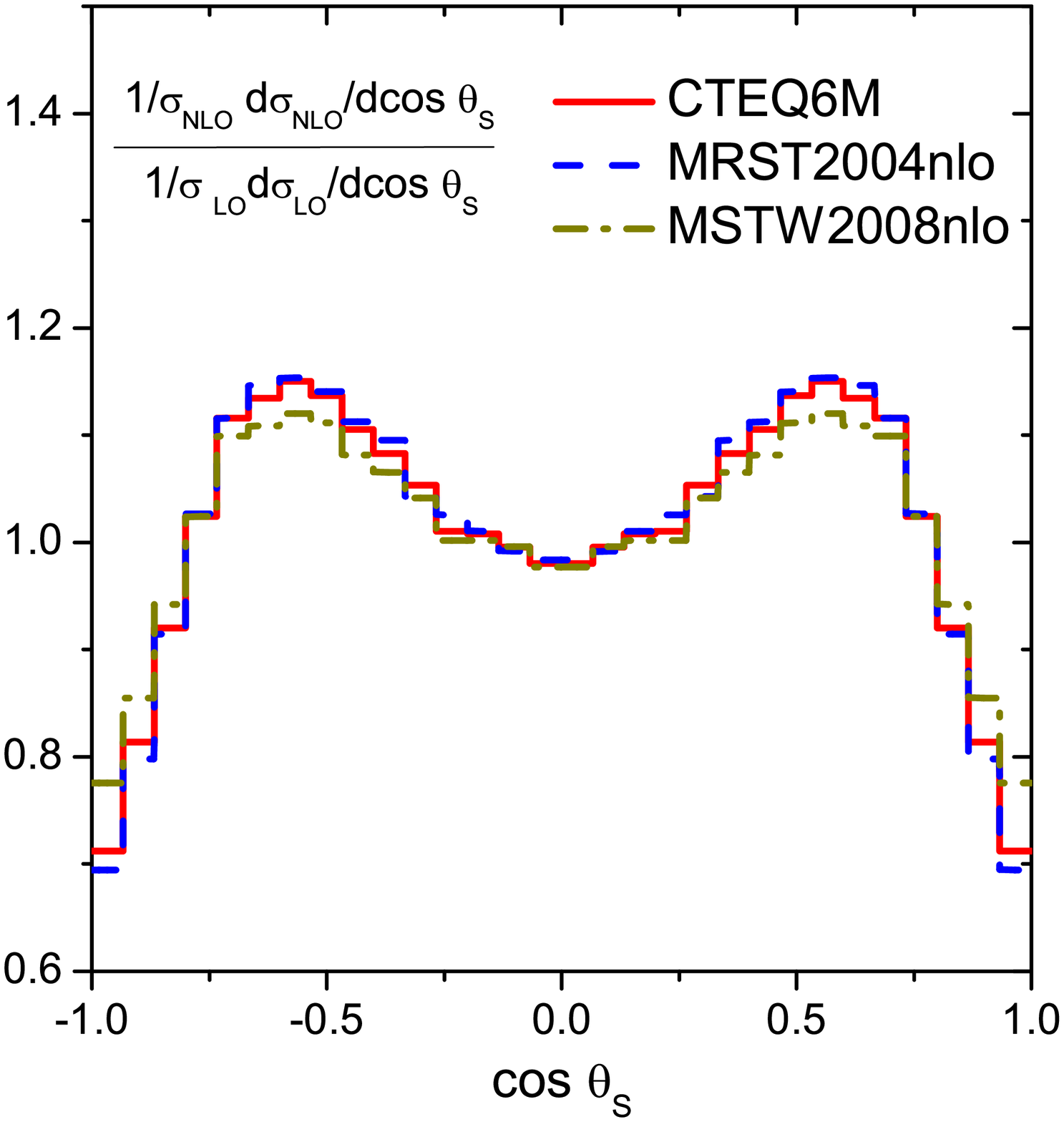}
\caption[]{The scale and PDF uncertainties of the NLO polar angle
distribution for the graviton at the LHC with $\sqrt s=14{\rm\ TeV}$
and $m_X=3000{\rm\ GeV}$.} \label{f8}
\end{figure}

In Figs.~\ref{f5}-\ref{f7} we show the normalized polar angle
distributions of the top quark at the LO and the ratios of the
normalized NLO distributions to the LO ones at the LHC with $\sqrt
s=14{\rm\ TeV}$. At the LO, we can use both the $\cos \theta_S$ and
$\cos \theta^*$ distributions to distinguish the $Z'$ and the $G$ as
their distributions have significantly different shapes. At the same
time we can also use the $\cos \theta^*$ distribution to distinguish
between the $Z'_{1,2}$ and the $Z'_{3,4}$ since the latter ones have
a large forward-backward asymmetry. Furthermore, the differences of
the polar angle distributions between the $Z'_1$ and the $Z'_2$ are
very small for $m_X$ around $1{\rm\ TeV}$ or heavier, thus it is not
possible to separate them through the polar angle distributions. We
present the LO squared helicity amplitudes in the Appendix, which can
explain the behavior of the LO polar angle distributions. After
including the NLO corrections, we can see that for all the $Z'$ the
changes of the distributions are negligibly small, which are no more
then a few percent. But for the $G$, as the increasing of the
resonance mass the NLO corrections can change the shapes of the
distributions, for example, the NLO corrections make the
distributions decrease more quickly at the both ends and can reach
about 10\% for $m_X=1500{\rm\ GeV}$. The corrections can be as large
as 30\%, and change the shapes of the distributions significantly for
$m_X=3000{\rm\ GeV}$, which do not change the fact that the
distributions of the $G$ and the $Z'$ are greatly different. Note
that for a heavy enough resonance, the top quarks produced are highly
boosted, which means the decay products of the top quark are close to
each other and form a top jet. Recently, several methods based on the
jet substructures have been proposed ~\cite{Kaplan:2008ie}, which may
be used to detect such a top jet efficiently, so it is possible to
measure the NLO QCD effects to the polar angle distributions for a
graviton with a mass of several ${\rm TeV}$ at the LHC. The reason
that the NLO distributions for the graviton differ from the LO ones
is that the NLO corrections change the ratio of the contributions
from  the $gg$ and $q\bar q$ channels, as shown in Sec.~\ref{tot},
which have different shapes of distributions. As the resonance mass
increases, the contributions from these two channels become
comparable, so the changes are more significant. We further study the
scale and PDF uncertainties of the NLO polar angle distributions for
a graviton with $m_X=3000\ {\rm GeV}$. As shown in Fig.~\ref{f8}, the
uncertainty from the scale dependence is negligibly small, and the
PDF uncertainty is within 10\%, which is still small as compared to
the NLO corrections. Here, we use two more PDF sets in the NLO
calculations, i.e., the MRST2004nlo~\cite{Martin:2004ir} and
MSTW2008nlo~\cite{Martin:2009bu} PDF sets.

\begin{figure}[h!]
\includegraphics[width=0.4\textwidth]{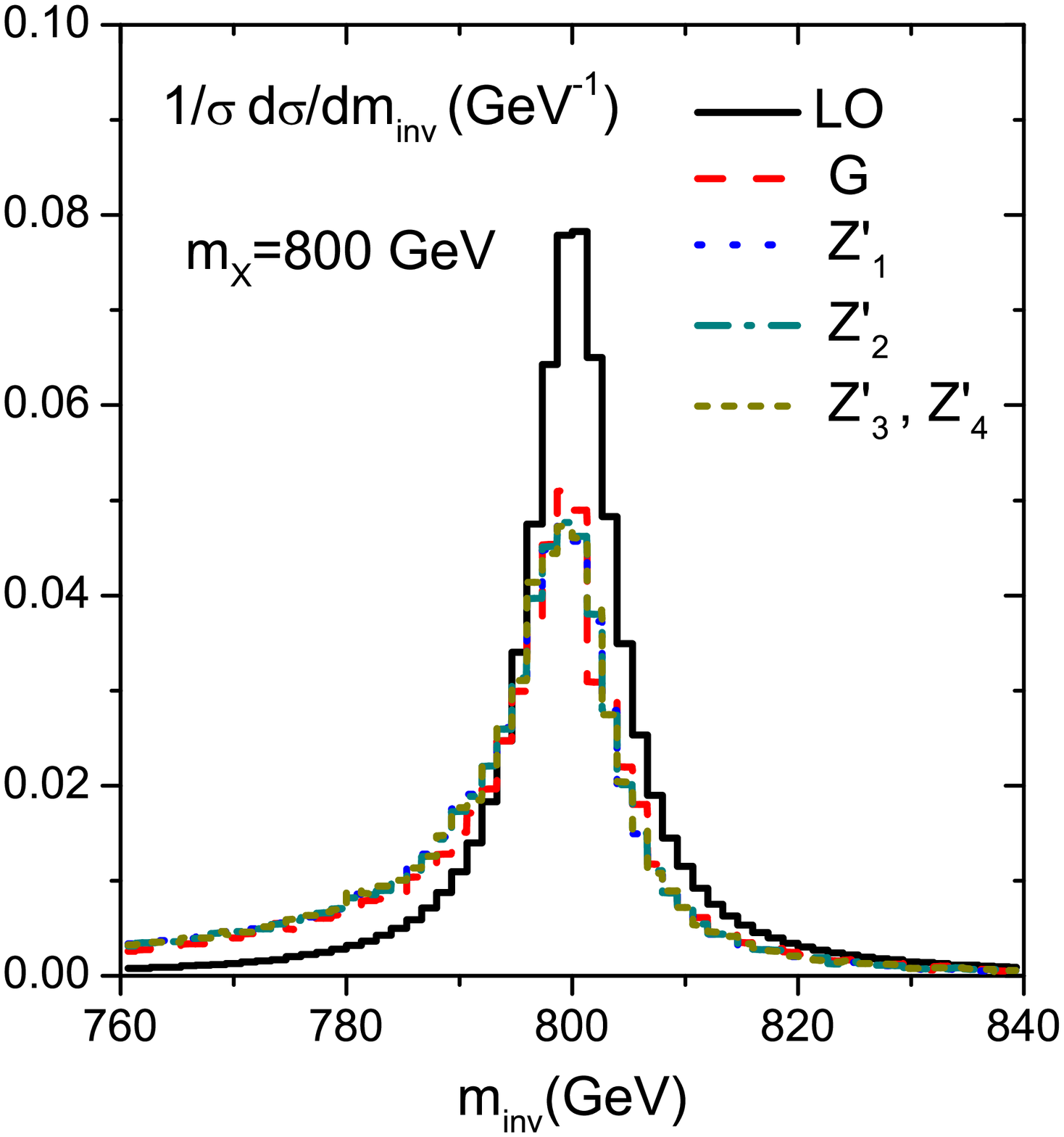}
\includegraphics[width=0.4\textwidth]{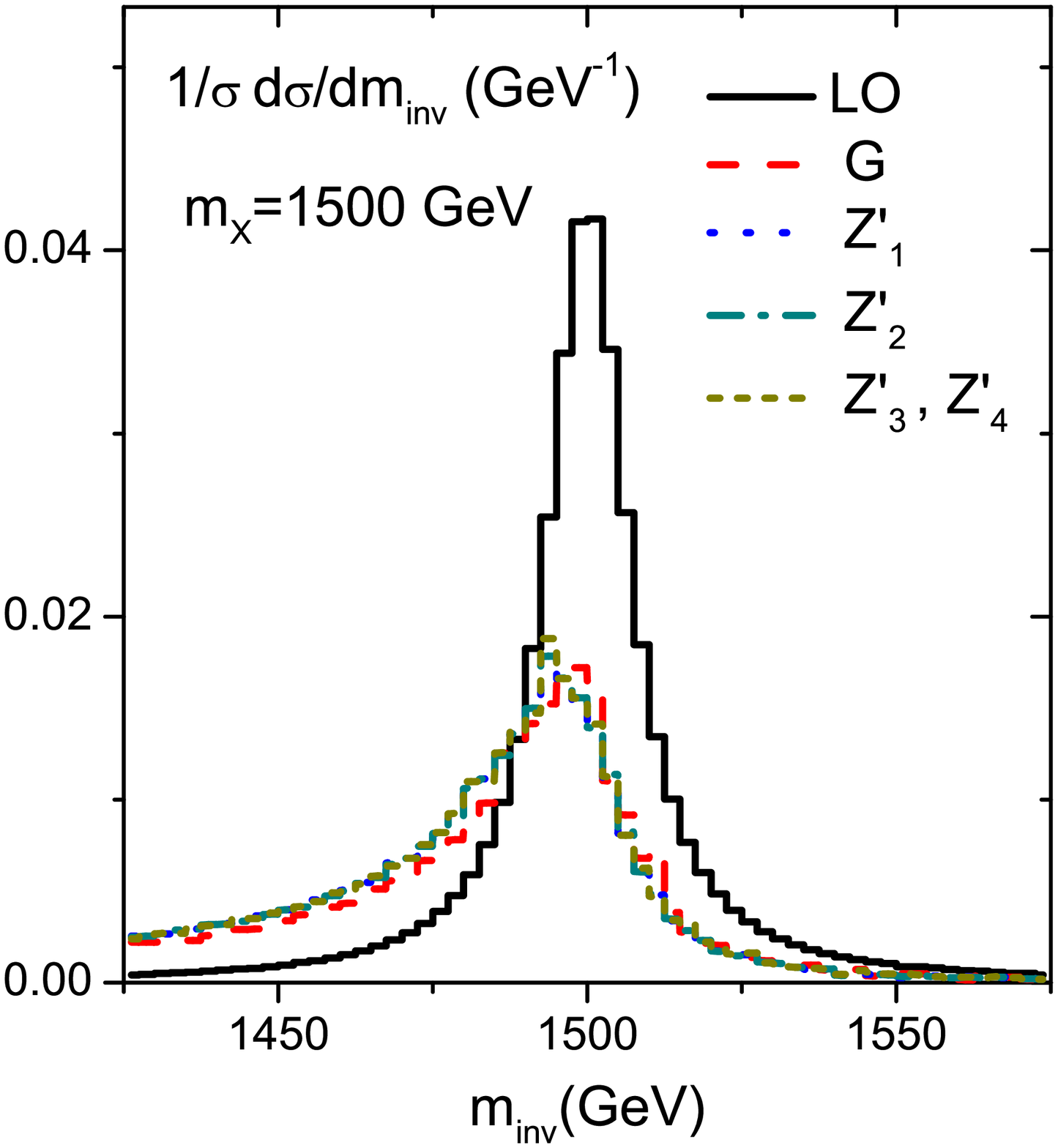}
\caption[]{The normalized top quark pair invariant mass distributions
at both the LO and the NLO at the LHC with $\sqrt s=14{\rm\ TeV}$.}
\label{f9}
\end{figure}

In Fig.~\ref{f9} we present the invariant mass distributions of the
top quark pair including the NLO QCD corrections. At the LO they are
just the Breit-Wigner distributions with a center value $m_X$ and a
width $\Gamma_X$. While at the NLO the heavy resonance can decay into
a top quark pair plus a hard gluon, so the NLO corrections increase
the distributions in the lower invariant mass value region, and the
changes of the distributions are more significant as the resonance
mass increases. We also studied all the above distributions at the
LHC with $\sqrt s= 7{\rm\ TeV}$, and the results are similar.

\subsection{The spin correlations}
One of the unique features of the top quark is that it decays before
the strong interaction can depolarize its spin. Thus, it is possible
to extract the spin information of the produced top quark by studying
the angular distributions of the decay products. For a spin up top
quark (or a spin down anti-top quark), the decay angular distribution
of the $i$th decay product is given by~\cite{Mahlon:2000ze}
\begin{equation}
{1\over \Gamma_T}\frac{d\Gamma}{d(\cos\chi_i)}={1\over
2}\left(1+\alpha_i\cos\chi_i\right),
\end{equation}
where $i$ could be quarks, $b, u, c, \bar{d}, \bar{s}$, or leptons,
$\nu_l, \bar{l}$; $\chi_i$ is the angle between the $i$th decay
product and the spin quantization axis in the top rest frame, and
$\alpha_i$ are the correlation coefficients. For the charged leptons,
$\alpha_l=1$ exactly, which means the charged leptons are maximally
correlated with the top spin direction.

The spin correlations of the top quark pair also can be used for the
identification of the heavy resonance, but the precision
measurement of them is more difficult at the LHC. In order to study
the spin correlations of the top quark pair production, the following
double differential cross section is usually considered,
\begin{equation}
{1\over \sigma}\frac{d^2\sigma}{d(\cos\chi_i^+)d(\cos\chi_j^-)}=
{1\over4}\left(1-A\alpha_i\alpha_j\cos\chi_i^+\cos\chi_j^-
+b_+\alpha_i\cos\chi_i^+ + b_-\alpha_j\cos\chi_j^-\right),
\end{equation}
neglecting the interference between the top spins we have
\begin{eqnarray}
A&=&\frac{\sigma(t_{\uparrow}\bar t_{\uparrow}+t_{\downarrow}\bar
t_{\downarrow})-\sigma(t_{\uparrow}\bar
t_{\downarrow}+t_{\downarrow}\bar
t_{\uparrow})}{\sigma(t_{\uparrow}\bar
t_{\uparrow}+t_{\downarrow}\bar
t_{\downarrow})+\sigma(t_{\uparrow}\bar
t_{\downarrow}+t_{\downarrow}\bar t_{\uparrow})}, \nonumber\\
b_+&=&\frac{\sigma(t_{\uparrow}\bar t_{\uparrow}+t_{\uparrow}\bar
t_{\downarrow})-\sigma(t_{\downarrow}\bar
t_{\uparrow}+t_{\downarrow}\bar
t_{\downarrow})}{\sigma(t_{\uparrow}\bar
t_{\uparrow}+t_{\downarrow}\bar
t_{\downarrow})+\sigma(t_{\uparrow}\bar
t_{\downarrow}+t_{\downarrow}\bar t_{\uparrow})}, \nonumber\\
b_-&=&\frac{\sigma(t_{\uparrow}\bar
t_{\downarrow}+t_{\downarrow}\bar
t_{\downarrow})-\sigma(t_{\uparrow}\bar
t_{\uparrow}+t_{\downarrow}\bar
t_{\uparrow})}{\sigma(t_{\uparrow}\bar
t_{\uparrow}+t_{\downarrow}\bar
t_{\downarrow})+\sigma(t_{\uparrow}\bar
t_{\downarrow}+t_{\downarrow}\bar t_{\uparrow})}.
\end{eqnarray}
In our following calculations we use the helicity basis in the
$t\bar t$ center of mass frame, which means $\chi_i^+(\chi_j^-)$ is
defined to be the angle between the $t(\bar t)$ direction in the
$t\bar t$ center of mass frame and the corresponding decay product
direction in the $t(\bar t)$ rest frame.

\begin{figure}[h!]
\includegraphics[width=0.7\textwidth]{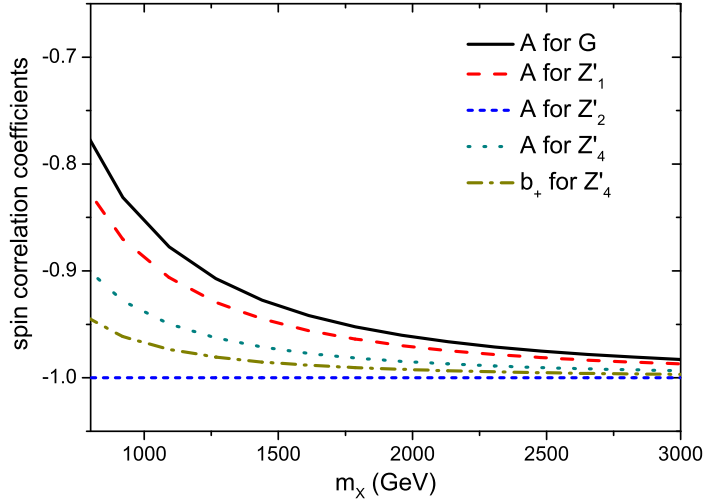}
\caption[]{The top quark pair spin correlation coefficients at the LO as
functions of the heavy resonance mass at the LHC with $\sqrt s=14{\rm
\ TeV}$.} \label{f10}
\end{figure}

In Fig.~\ref{f10} we show the LO results of the top quark pair spin
correlation coefficients as functions of the heavy resonance mass at
the LHC with $\sqrt s=14{\rm\ TeV}$. According to symmetry analysis
we have
\begin{eqnarray}
&&A(Z'_3)=A(Z'_4), \quad b_{\pm}(G,Z'_1,Z'_2)=0,\nonumber \\
&&b_+(Z'_3)=b_-(Z'_3)=-b_+(Z'_4)=-b_-(Z'_4),
\end{eqnarray}
and at the LO for the axial vector $Z'_2$,
\begin{equation}
A(Z'_2)=-1,
\end{equation}
which can be seen from the helicity amplitudes in the Appendix. With
the increasing heavy resonance mass, all the coefficients in
Fig.~\ref{f10} approach $-1$ due to the fact that the cross sections
for the $t\bar t$ with the same helicities vanish as the heavy
resonance mass goes infinity. In Table~\ref{t1}, we list some typical
NLO results of those coefficients. We can see that the NLO QCD
corrections are rather small, about $1\%-2\%$, and can be neglected
at the LHC. We also investigate the cases for $\sqrt s=7{\rm\ TeV}$,
and the results are almost the same at both the LO and the NLO.

\begin{table*}
\begin{center}
\scalebox{1}[0.9] {\begin{tabular}{|c|c|cc|cc|cc|} \hline $\
$mass(GeV)$\ $& $\ $resonance$\ $ &$\ $$A$(LO)$\ $&$A$(NLO)$\ $&$\
$$b_+$(LO)$\ $&$b_+$(NLO)$\ $&$\ $$b_-$(LO)$\ $&$b_-$(NLO)$\ $
\\
\hline \multirow{5}{*}{800}&$G$&-0.783&-0.778&0&0&0&0
\\
&$Z'_1$&-0.832&-0.826&0&0&0&0
\\
&$Z'_2$&-1.000&-0.986&0&0&0&0
\\
&$Z'_3$&-0.904&-0.896&0.947&0.943&0.947&0.943
\\
&$Z'_4$&-0.904&-0.896&-0.947&-0.943&-0.947&-0.943
\\
\hline \multirow{5}{*}{1500}&$G$&-0.933&-0.913&0&0&0&0
\\
&$Z'_1$&-0.949&-0.933&0&0&0&0
\\
&$Z'_2$&-1.000&-0.980&0&0&0&0
\\
&$Z'_3$&-0.974&-0.955&0.986&0.977&0.986&0.977
\\
&$Z'_4$&-0.974&-0.955&-0.986&-0.977&-0.986&-0.977
\\
\hline
\end{tabular}}
\end{center}
\caption{The top quark pair spin correlation results at the LHC with
$\sqrt s=14{\rm\ TeV}$.} \label{t1}
\end{table*}

\section{conclusions}\label{s5}
We have calculated the complete NLO QCD corrections to a heavy
resonance production and decay into a top quark pair at the LHC,
where the resonance could be either a RS KK graviton $G$ or an extra
gauge boson $Z'$. Our results show that the total NLO K factors can
reach about $1.8- 2.0$ and $1.2- 1.4$ for the $G$ and all four types
of $Z'$ bosons, respectively, depending on the resonance mass. And
the NLO corrections from the production part are dominant, while the
ones from the decay part are relatively small but can still reach
above ten percent in some parameter regions. We also explore in
detail the NLO corrections to the polar angle distributions of the
top quark, and our results show that the NLO distributions are almost
the same as the LO ones for all four types of $Z'$ bosons, while the
shapes of the NLO distributions can be significantly different from
the LO ones for the $G$, depending on the mass of the resonance.
Moreover, the NLO corrections can also change the shapes of the top
quark pair invariant mass distributions. Finally, we study the NLO
corrections to the spin correlations of the top quark pair, and find
that the corrections are negligibly small.

\begin{acknowledgments}
This work was supported in part by the National Natural Science
Foundation of China, under Grants No.10721063, No.10975004 and
No.10635030. C.P.Y was supported in part by the U.S. National Science
Foundation under Grand No. PHY-0855561.
\end{acknowledgments}

\newpage

\section*{Appendix}
In this appendix we give the individual nonvanishing LO squared
helicity amplitudes for a heavy resonance production and decay into
a top quark pair, $q\bar q\ (gg)\rightarrow X \rightarrow t\bar t$. For
all the $Z'$ mediated processes,
\begin{eqnarray}
\overline{|\mathcal{M}|^2}_{Z'_1}&=&\left\{\begin{array}{l}
(1-\beta^2)\sin^2(\theta){\mathcal A}, \ \ {\rm for\ helicities\ \{+-++\},\ \{+---\}},\\
\hspace{3.8cm} {\rm \{-+++\}\ and \ \{-+--\}} \\
4\sin^4(\theta/2){\mathcal A}, \hspace{1cm} {\rm for\ helicities\ \{+--+\}\ and\ \{-++-\}}\\
4\cos^4(\theta/2){\mathcal A}, \hspace{1cm} {\rm for\ helicities\ \{+-+-\}\ and\ \{-+-+\}}, \\
\end{array}\right.\\
\overline{|\mathcal{M}|^2}_{Z'_2}&=&\left\{\begin{array}{l}
4\beta^2\sin^4(\theta/2){\mathcal A}, \hspace{0.8cm} {\rm for\ helicities\ \{+--+\}\ and\ \{-++-\}}\\
4\beta^2\cos^4(\theta/2){\mathcal A}, \hspace{0.8cm} {\rm for\ helicities\ \{+-+-\}\ and\ \{-+-+\}}, \\
\end{array}\right.\\
\overline{|\mathcal{M}|^2}_{Z'_3}&=&\left\{\begin{array}{l}
(1-\beta^2)\sin^2(\theta){\mathcal A}/4, \ \ {\rm for\ helicities\ \{+-++\}\ and\ \{+---\}}\\
(1-\beta)^2\sin^4(\theta/2){\mathcal A}, \ \ {\rm for\ helicities\ \{+--+\}}\\
(1+\beta)^2\cos^4(\theta/2){\mathcal A}, \ \ {\rm for\ helicities\ \{+-+-\}}, \\
\end{array}\right.\\
\overline{|\mathcal{M}|^2}_{Z'_4}&=&\left\{\begin{array}{l}
(1-\beta^2)\sin^2(\theta){\mathcal A}/4, \ \ {\rm for\ helicities\ \{-+++\}\ and\ \{-+--\}}\\
(1-\beta)^2\sin^4(\theta/2){\mathcal A}, \ \ {\rm for\ helicities\ \{-++-\}}\\
(1+\beta)^2\cos^4(\theta/2){\mathcal A}, \ \ {\rm for\ helicities\ \{-+-+\}}, \\
\end{array}\right.
\end{eqnarray}
and for the graviton mediated processes through the $q\bar q$
annihilation and the $gg$ fusion,
\begin{eqnarray}
\overline{|\mathcal{M}|^2}_{G,q\bar q}&=&\left\{\begin{array}{l}
\beta^2(1-\beta^2)\sin^2(2\theta){\mathcal B}/64, \ \ {\rm for\ helicities\ \{+-++\},\ \{+---\}}, \\
\hspace{5cm} {\rm \{-+++\}\ and \ \{-+--\}} \\
\beta^2(1+2\cos(\theta))^2\sin^4(\theta/2){\mathcal B}/16, \ \ {\rm for\ helicities\ \{+--+\}},\\
\hspace{6.5cm} {\rm and\ \{-++-\}}\\
\beta^2(1-2\cos(\theta))^2\cos^4(\theta/2){\mathcal B}/16, \ \ {\rm for\ helicities\ \{+-+-\}},\\
\hspace{6.5cm} {\rm and\ \{-+-+\}},\\
\end{array}\right.\\
\overline{|\mathcal{M}|^2}_{G,gg}&=&\left\{\begin{array}{l}
3\beta^2(1-\beta^2)\sin^4(\theta){\mathcal B}/128, \ \ {\rm for\ helicities\ \{+-++\},\ \{+---\}}, \\
\hspace{5cm} {\rm \{-+++\}\ and \ \{-+--\}} \\
3\beta^2\sin^6(\theta/2)\cos^2(\theta/2){\mathcal B}/8, \ \ {\rm for\ helicities\ \{+--+\},\ and\ \{-++-\}} \\
3\beta^2\sin^2(\theta/2)\cos^6(\theta/2){\mathcal B}/8, \ \ {\rm for\ helicities\ \{+-+-\},\ and\ \{-+-+\}}, \\
\end{array}\right.
\end{eqnarray}
with
\begin{equation}
{\mathcal A}=\frac{s^2}{(s-m_{Z'}^2)^2+m_{Z'}^2\Gamma_{Z'}^2},\ \
{\mathcal B}=\frac{s^4}{(s-m_G^2)^2+m_G^2\Gamma_G^2},
\end{equation}
where $\theta$ is the polar angle between the momenta of the top
quark and the light quark (or gluon) in the center of the mass frame
of the top quark pair, $s$ is the square of the center of mass
energy, $\beta \equiv \sqrt{1-4m_{top}^2/s}$, and the squared
amplitudes have been summed and averaged over the color of the
external particles.


\begin{thebibliography}{999}
\bibitem{Hill:2002ap}
  C.~T.~Hill and E.~H.~Simmons,
  Phys.\ Rept.\  {\bf 381} (2003) 235
  [Erratum-ibid.\  {\bf 390} (2004) 553].

\bibitem{Hill:1991at}
  C.~T.~Hill,
  Phys.\ Lett.\  B {\bf 266} (1991) 419;
  C.~T.~Hill and S.~J.~Parke,
  Phys.\ Rev.\  D {\bf 49} (1994) 4454.

\bibitem{ArkaniHamed:2001nc}
  N.~Arkani-Hamed, A.~G.~Cohen and H.~Georgi,
  Phys.\ Lett.\  B {\bf 513} (2001) 232;
  M.~Schmaltz and D.~Tucker-Smith,
  Ann.\ Rev.\ Nucl.\ Part.\ Sci.\  {\bf 55} (2005) 229.

\bibitem{Langacker:2008yv}
  P.~Langacker,
  Rev.\ Mod.\ Phys.\  {\bf 81} (2008) 1199.

\bibitem{Godfrey:2008vf}
  S.~Godfrey and T.~A.~W.~Martin,
  Phys.\ Rev.\ Lett.\  {\bf 101} (2008) 151803.

\bibitem{Randall:1999ee}
  L.~Randall and R.~Sundrum,
  Phys.\ Rev.\ Lett.\  {\bf 83} (1999) 3370.

\bibitem{Fitzpatrick:2007qr}
  A.~L.~Fitzpatrick, J.~Kaplan, L.~Randall and L.~T.~Wang,
  JHEP {\bf 0709} (2007) 013.

\bibitem{Gherghetta:2000qt}
  T.~Gherghetta and A.~Pomarol,
  Nucl.\ Phys.\  B {\bf 586} (2000) 141;
  B.~Lillie, L.~Randall and L.~T.~Wang,
  JHEP {\bf 0709} (2007) 074;
  B.~Lillie, J.~Shu and T.~M.~P.~Tait,
  Phys.\ Rev.\  D {\bf 76} (2007) 115016;
  U.~Baur and L.~H.~Orr,
  Phys.\ Rev.\  D {\bf 77} (2008) 114001.


\bibitem{Agashe:2007ki}
  K.~Agashe {\it et al.},
  Phys.\ Rev.\  D {\bf 76} (2007) 115015;
  A.~Djouadi, G.~Moreau and R.~K.~Singh,
  Nucl.\ Phys.\  B {\bf 797} (2008) 1.

\bibitem{Barger:2006hm}
  V.~Barger, T.~Han and D.~G.~E.~Walker,
  Phys.\ Rev.\ Lett.\  {\bf 100} (2008) 031801.
\bibitem{Frederix:2007gi}
  R.~Frederix and F.~Maltoni,
  JHEP {\bf 0901} (2009) 047;
  Y.~Bai and Z.~Han,
  JHEP {\bf 0904} (2009) 056.

\bibitem{Mathews:2005bw}
  P.~Mathews, V.~Ravindran and K.~Sridhar,
  JHEP {\bf 0510} (2005) 031.

\bibitem{Li:2006yv}
  Q.~Li, C.~S.~Li and L.~L.~Yang,
  Phys.\ Rev.\  D {\bf 74} (2006) 056002.

\bibitem{Ball:2007zza}
  G.~L.~Bayatian {\it et al.}  [CMS Collaboration],
  J.\ Phys.\ G {\bf 34} (2007) 995.

\bibitem{Csaki:2004ay}
  C.~Csaki,
  arXiv:hep-ph/0404096.

\bibitem{Han:1998sg}
  T.~Han, J.~D.~Lykken and R.~J.~Zhang,
  Phys.\ Rev.\  D {\bf 59} (1999) 105006.

\bibitem{Campbell:2004ch}
  J.~M.~Campbell, R.~K.~Ellis and F.~Tramontano,
  Phys.\ Rev.\  D {\bf 70} (2004) 094012;
  Q.~H.~Cao and C.~P.~Yuan,
  Phys.\ Rev.\  D {\bf 71} (2005) 054022.

\bibitem{'tHooft:1972fi}
  G.~'t Hooft and M.~J.~G.~Veltman,
  Nucl.\ Phys.\  B {\bf 44} (1972) 189.

\bibitem{Chanowitz:1979zu}
  M.~S.~Chanowitz, M.~Furman and I.~Hinchliffe,
  Nucl.\ Phys.\  B {\bf 159} (1979) 225.

\bibitem{Harris:2001sx}
  B.~W.~Harris and J.~F.~Owens,
  Phys.\ Rev.\  D {\bf 65} (2002) 094032.

\bibitem{Amsler:2008zzb}
  C.~Amsler {\it et al.}  [Particle Data Group],
  Phys.\ Lett.\  B {\bf 667} (2008) 1.

\bibitem{Pumplin:2002vw}
  J.~Pumplin, D.~R.~Stump, J.~Huston, H.~L.~Lai, P.~M.~Nadolsky and W.~K.~Tung,
  JHEP {\bf 0207} (2002) 012.

\bibitem{Collins:1977iv}
  J.~C.~Collins and D.~E.~Soper,
  Phys.\ Rev.\  D {\bf 16} (1977) 2219.

\bibitem{Kaplan:2008ie}
  D.~E.~Kaplan, K.~Rehermann, M.~D.~Schwartz and B.~Tweedie,
  Phys.\ Rev.\ Lett.\  {\bf 101} (2008) 142001;
  J.~Thaler and L.~T.~Wang,
  JHEP {\bf 0807} (2008) 092;
  S.~D.~Ellis, C.~K.~Vermilion and J.~R.~Walsh,
  Phys.\ Rev.\  D {\bf 80} (2009) 051501;
  G.~Giurgiu  [for the CMS collaboration],
  arXiv:0909.4894 [hep-ex].

\bibitem{Martin:2004ir}
  A.~D.~Martin, R.~G.~Roberts, W.~J.~Stirling and R.~S.~Thorne,
  Phys.\ Lett.\  B {\bf 604} (2004) 61.

\bibitem{Martin:2009bu}
  A.~D.~Martin, W.~J.~Stirling, R.~S.~Thorne and G.~Watt,
  Eur.\ Phys.\ J.\  C {\bf 64} (2009) 653.

\bibitem{Mahlon:2000ze}
  G.~Mahlon,
  arXiv:hep-ph/0011349.


\end{thebibliography}
\end{document}